\documentclass[twocolumn,showpacs,pre,eqsecnum,citeautoscript,amsmath,amssymb,floatfix,superscriptaddress]{revtex4-1}
\usepackage{bm,color,amsmath,amssymb,mathrsfs,latexsym,graphicx,psfrag,float}

\newcommand{\opt}[1]

\usepackage{comment}
\usepackage{hyperref}
\usepackage{enumerate}
\usepackage{txfonts}
\usepackage{bbm}
\usepackage{ifthen}
\usepackage{cleveref}
\usepackage{dsfont}
\usepackage{tabularx}
\newcolumntype{L}[1]{>{\raggedright\arraybackslash}p{#1}} 
\newcolumntype{C}[1]{>{\centering\arraybackslash}p{#1}} 
\newcolumntype{R}[1]{>{\raggedleft\arraybackslash}p{#1}} 

\newcommand{\eq}[2]{\begin{align}\label{#1} #2 \end{align}}

\usepackage{color}

\begin{document}

\title{Background Field Functional Renormalization Group for Absorbing State Phase Transitions}

\author{Michael Buchhold}
\affiliation{Institut f\"ur Theoretische Physik, Universit\"at zu K\"oln, D-50937 Cologne, Germany}
\author{Sebastian Diehl}
\affiliation{Institut f\"ur Theoretische Physik, Universit\"at zu K\"oln, D-50937 Cologne, Germany}

\begin{abstract}
We present a functional renormalization group approach for the active to inactive phase transition in directed percolation type systems, in which the transition is approached from the active, finite density phase. By expanding the effective potential for the density field around its minimum, we obtain a background field action functional, which serves as a starting point for the functional renormalization group approach. Due to the presence of the background field, the corresponding non-perturbative flow equations yield remarkably good estimates for the critical exponents of the directed percolation universality class, even in low dimensions.

\end{abstract}


\maketitle
\section{Introduction}
The dynamics of a large class of important physical systems features a so-called absorbing state, i.e. a state which can be accessed during the time evolution of the system but once reached, it can never be left \cite{NEQ_PT1, DP_Hinrichsen, Janssen1981, Tauber-book}. An intrinsic property of such an absorbing state dynamics is the violation of detailed balance, and therefore the absence of an equilibrium-type partition function that can be expressed in terms of a classical Hamiltonian functional alone \cite{Tauber-book, Kamenev}. Typically, these models feature a second order phase transition from an active phase, for which the stationary state is not the absorbing state, to an inactive phase with absorbing stationary state. The absorbing state phase transitions correspond to genuine non-equilibrium universality classes without equilibrium counterpart, and their characterization in terms of symmetries and critical exponents is a major issue of non-equilibrium statistical mechanics \cite{Elgart2006, Kamenev, DP_Hinrichsen, Tauber-book, Cardy1996}. 

Many known absorbing state phase transitions belong to the realm of classical non-equilibrium statistical mechanics and have been investigated via perturbative \cite{Janssen2005, Tauber2005, Bronzan1974, Janssen1981, Garrahan2005} and non-perturbative \cite{Canet2004, Canet2005, Canet2006, Basu2011, Gredat2014} renormalization group techniques and efficient numerical routines for classical dynamical systems \cite{Jensen1999, Voigt1997, Jensen1992}. However, only recently there have been proposals to experimentally realize cold atomic Rydberg systems, which feature absorbing state phase transitions in the presence of competing classical and quantum dynamics \cite{Marcuzzi2015, Mattioli2015, Marcuzzi2016}. Since numerical simulations of quantum systems, especially with fermion or spin statistics, are much more demanding than for classical systems, it is important to develop adequate reliable analytical methods for the classification of these new absorbing state phase transitions. 

In this work, we introduce a background field functional renormalization group (FRG) approach \cite{Dashen1981,Wetterich1993,Berges2002,Schoeller2007,Ellwanger,Litim2002,Litim2002a} for absorbing state phase transitions, which represents a non-perturbative renormalization group approach to the phase transition with a microscopic starting point in the active phase. In contrast to previous approaches to absorbing state phase transitions \cite{Canet2004, Canet2005, Canet2006, Janssen2005, Tauber2005} this approach explicitly breaks the characteristic symmetry of the underlying model during the renormalization group flow but is tailored to preserve the scaling behavior of the fields, which is characteristic for directed percolation systems.
The strength of this approach is the effective inclusion of diagrams beyond one-loop order, which lead to an effective field dependence of the wave function renormalization and diffusion constant, combined with the well-known non-perturbative nature of the functional renormalization group. The physical interpretation of this symmetry breaking is that the corresponding noise level at the critical point is strongly increased compared to the deterministic part of the action.
We show that this approach yields analytic and remarkably precise expressions for the universal exponents of the absorbing state transition.
On the other hand, as a consequence of the explicit symmetry breaking, this approach is only reliable at the fixed point, where scaling behavior is found, and not suited to determine non-universal properties such as the phase boundary or the order parameter expectation value.
While this method can be generally applied to absorbing state phase transitions, we demonstrate its applicability and the quality of its results on a specific class of absorbing state systems, which is directed percolation (DP) \cite{DP_Hinrichsen}. Directed percolation represents the most common and best understood class of absorbing state phase transitions and serves as a benchmark system for the present renormalization group approach. However, we expect this approach to work for a large class of absorbing state transitions, for which an imbalance between the noisy and the deterministic part of the action can be implemented.

This paper is organized as follows. In Sec.~\ref{sec:DP} we introduce the directed percolation dynamics and review basic properties of the corresponding systems. In Sec.~\ref{sec:FRG}, we introduce the background field functional renormalization group approach in the active phase and subsequently apply it to the absorbing state phase transition of the DP model in Sec.~\ref{sec:Exp}. Finally, we conclude this work in Sec.~\ref{sec:Conc}.

\section{Directed Percolation}\label{sec:DP}
The critical dynamics of a large class of non-equilibrium models is described by the directed percolation universality class. Among these are contact
processes \cite{Harris1974}, Reggeon Field Theory \cite{Reggeon, Moshe1978, Grassberger1978},  autocatalytic reaction models
\cite{Sch1972, Janssen1981, Aukrust1990}, supercooled liquids \cite{Garrahan2004,Garrahan2005}, and the recently proposed ultracold, driven Rydberg setups \cite{Ryd-int, PRL-KinC, Marcuzzi2015}. This makes such dynamics not only interesting from a theoretical standpoint, but also in view of possible near future experiments with flexible dimensionality \cite{DP_exp, DP_explong}. The dynamics close to the absorbing state transition is described by the well-known DP action 
\eq{action1}{
S=\int_X \left[\tilde{\varphi}\left(\partial_t-D\nabla^2-\Delta\right)\varphi+\mu\tilde{\varphi}\varphi(\varphi-\tilde{\varphi})\right].
}
Here, $\varphi$ is a real scalar field in $d+1$ spatial and temporal dimensions, which represents a density of active sites, $\tilde{\varphi}$ represents a purely imaginary scalar response field, $\int_X=\int_{x,t}$ the integral over time and space, $D$ the diffusion constant, $\Delta$ the gap and $\mu$ the strength of the non-linearity. Due to the imaginary nature of the response field, the partition function is normalized to one, i.e.
\eq{action1p5}{
1=\mathcal{Z}=\int\mathcal{D}[\tilde{\varphi},\varphi]e^{-S}.
}
In order to prepare the discussion in the following sections, we use the remainder of this section to review the basic properties of the action \eqref{action1}, which are detailed in a number of excellent reviews and textbooks \cite{DP_Hinrichsen, Kamenev, Altland}.

One very important feature of the DP action \eqref{action1} (and more generally, absorbing state problems) is the absence of an additive noise scale, i.e.  of a static fluctuation term $\sim T\tilde{\varphi}^2$, and implies the mentioned absence of detailed balance. The physical origin behind this property is that the absorbing state, for which the density $\varphi=0$ vanishes, is completely fluctuationless. This is made obvious by deriving the corresponding Langevin equation from Eq.~\eqref{action1}, which reads
\eq{action2}{
\partial_t\varphi=(D\nabla^2+\Delta)\varphi-\mu\varphi^2+\xi,
}
with a Markovian multiplicative noise $\xi$, with vanishing average $\langle\xi\rangle=0$ and variance
\eq{action3}{
\langle\xi_X\xi_{X'}\rangle=2\delta(X-X')\mu\varphi_X.
}
Consequently, the noise is multiplicative $\xi\sim\sqrt{\varphi}$ and therefore vanishes completely for $\varphi\rightarrow0$, which is different from equilibrium models, where the temperature $T$ serves as the minimum noise scale.\\
On a formal level, the structure of the action implies that $S_{\tilde{\varphi}=0}=0$ as well as $S_{\varphi=0}=0$. While the former relation is generic for equilibrium and non-equilibrium models and is enforced by causality, the latter is specific to non-equilibrium models with multiplicative noise fluctuations.

In the present model, the absence of a static noise term is further guaranteed by the characteristic symmetry of directed percolation, which is called rapidity inversion symmetry \cite{Janssen2005, Kamenev}. It is expressed by the invariance of $S$ under the symmetry transformation
\eq{action4}{
\tilde{\varphi}\leftrightarrow-\varphi, \ \ \ \ t\rightarrow-t.
}
Together with the causality condition $S_{\tilde{\varphi}=0}=0$, this symmetry guarantees $S_{\varphi=0}=0$ also under renormalization and therefore the absence of a static noise $\sim T\tilde{\varphi}^2$ in the effective action $\Gamma$, including all fluctuations. Furthermore, it fixes the scaling of the fields $\tilde{\varphi}\sim \varphi$ to be equal.

We close this section by reviewing the canonical scaling analysis of the action \eqref{action1}, and discuss the corresponding mean-field exponents. In order to determine the canonical scaling of the constituents of the action, time and space are rescaled according to $x\rightarrow b x$, $t\rightarrow b^z t$. Here, $z$ is the dynamical critical exponent, which determines the relation of time and space according to $\omega\sim q^{z}$. Performing this rescaling, one finds
\eq{action5}{
S=\int_X b^{d+z}\left[\tilde{\varphi}\left(b^{-z}\partial_t-b^{-2}D\nabla^2-\Delta\right)\varphi+\mu\tilde{\varphi}\varphi\left(\varphi-\tilde{\varphi}\right)\right].
}
This sets $z=2$ on the mean-field level. In order to leave the fluctuating part of the action invariant under this transformation, the fields $\tilde{\varphi}, \varphi$ have to transform as $\tilde{\varphi}\varphi\rightarrow\tilde{\varphi}\varphi b^{-d}$. Because of the symmetry \eqref{action4}, this leads to
\eq{action6}{
\varphi\rightarrow b^{-d/2}\varphi, \ \  \ \tilde{\varphi}\rightarrow b^{-d/2}\tilde{\varphi}, \ \ \ \Delta\rightarrow \Delta b^{2}, \ \ \  \mu\rightarrow b^{2-\frac{d}{2}}\mu.
}
Under coarse graining, i.e. for $b>1$, the gap $\Delta$ is effectively growing. The same is true for the non-linearity $\mu$ in dimensions $d<4$, which introduces infrared divergent contributions from these terms at the critical point. On the other hand, for $d>4$, $\mu$ scales to zero and the effect of the non-linearity is only determined by its ultraviolet behavior, which defines $d_c=4$ as the upper critical dimension of DP. For $d>4$, the canonical scaling of Eqs.~\eqref{action5}, \eqref{action6} is not modified by fluctuation corrections, and one can directly determine the relevant critical exponents. These are the exponent $z$, determined above to be $z=2$, the scaling exponent of the order parameter $\langle\varphi\rangle\sim \Delta^{\beta}$, which is $\beta=1$ according to $\langle\varphi\rangle=\frac{\Delta}{\mu}$, and the scaling exponent of the correlation length $\xi$, which is $\nu=\frac{1}{2}$ according to $\xi=\sqrt{\frac{D}{\Delta}}=\Delta^{-\nu}$. 

In principle, one can add to the action \eqref{action1} any combination of fields that respects the symmetry \eqref{action4} and causality. The most relevant of such additional terms are the quartic vertices $ \nu_4\tilde{\varphi}\varphi(\varphi-\tilde{\varphi})^2$ and $\mu_4\varphi^2\tilde{\varphi}^2$, which are generated from \eqref{action1} already at one-loop order. Although the canonical power counting suggests that $\mu_4, \nu_4$ are less relevant than the leading order non-linearity $\mu$, they will be non-zero at the Wilson-Fisher fixed point in $d<4$ and therefore modify the critical exponents and may not be neglected. In Refs.~\cite{Canet2004, Canet2005}, the authors have applied functional renormalization group methods to absorbing state phase transitions and have included higher order vertices in their renormalization group scheme via a general potential of the invariants $x=\tilde{\varphi}\varphi, y=\varphi-\tilde{\varphi}$, obtaining good estimates for the critical exponents. In the following section, we will discuss a functional renormalization group approach in the finite density phase, which partially considers the effect of higher order vertices and which yields remarkably good critical exponents already on the basis of the minimal truncation \eqref{action1} evaluated with a background field approach~\cite{Litim2002,Litim2002a}.

\section{Background Field Functional Renormalization Group}\label{sec:FRG}
In this section, we discuss the functional integral approach to determine the partition function of the DP system in the active phase and introduce the functional renormalization group via the Wetterich equation \cite{Wetterich1993, Berges2002}. The latter represents a method to compute the effective action $\Gamma$, i.e. the generator of the one-particle irreducible response and correlation functions, starting from the microscopic action.

In order to obtain the low frequency action in the active phase, we expand the action around its classical saddle point \cite{Dashen1981}. The saddle point is determined by the equations
\eq{Saddle}{
\frac{\delta S}{\delta\tilde{\varphi}}=\frac{\delta S}{\delta\varphi}=0.
}
For $\Delta>0$, the saddle point density is different from zero $\varphi_0>0$, and the system is in the active phase. The corresponding solution of the saddle point equations is $\tilde{\varphi}_0=0, \varphi_0=\frac{\Delta}{\mu}$. Expanding the action around the saddle point solution yields
\eq{Saddle2}{
S=\int_X \tilde{\varphi}\left(\partial_t-D\nabla^2+\rho\right)\varphi+\mu\tilde{\varphi}\varphi\left(\varphi-\tilde{\varphi}\right)-\rho\tilde{\varphi}^2,
}
where we replaced $\rho=\mu\varphi_0$. This action has a positive gap $\rho>0$, which also appears as a non-zero noise term $\rho\tilde{\varphi}^2$. The latter is the consequence of the multiplicative noise in the Langevin equation \eqref{action2}, which transforms
\eq{Saddle3}{
\langle \xi_X\xi_{X'}\rangle=2\delta(X-X') \mu\varphi_X\rightarrow2\delta(X-X')\left(\mu\varphi_X+\rho\right)}
after expansion around the saddle point solution. 

The action \eqref{Saddle2} does no longer display the rapidity inversion symmetry \eqref{action4} and instead is invariant under the modified field transformation
\begin{eqnarray}\label{modsym}
\tilde{\varphi}\leftrightarrow-(\varphi+\rho/\mu),\ \ \ \ t\rightarrow-t,
\end{eqnarray}
which formally replaces the rapidity inversion symmetry \eqref{action4} in the active phase. It is, however, important to note that, in contrast to Eqs.~\eqref{action1}, \eqref{action4}, the modified field transformation connects vertices of different order in the fields, as it represents no longer a homogeneous transformation.
This is a consequence of the fact that the response field must not acquire a finite expectation value, guaranteed by causality. While the action \eqref{Saddle2} shows the modified rapidity inversion symmetry  (MRI), this symmetry is generally not preserved under renormalization group transformations, which are performed with a finite truncation in the vertices. Neglecting the generation of an $(n+1)$-body vertex during the RG flow will, due to the inhomogeneous nature of the MRI, automatically violate the MRI on the level of the $n$-body vertex, as one immediately realizes when writing down the corresponding flow equations.
A way to implement the MRI in the active phase is to compute the renormalization group flow of the complete, symmetrized and local potential $V(\tilde{\varphi},\varphi)$ of the fields as has been done in Refs.~\cite{Canet2004, Canet2005}, however for a form of the potential that respects the rapidity inversion symmetry. This approach is termed the local potential approximation and enforces a numerical evaluation of the flow equations. In the present approach, we accept the violation of the rapidity inversion symmetry during the flow but enforce the restoration of one major consequence of the symmetry, namely that both fields $\tilde{\varphi}\sim\varphi$ have exactly the same scaling behavior precisely at the fixed point, as we will further discuss below.

In order to determine the effective action $\Gamma$, which is identified as the generator of the one-particle irreducible correlation functions, we apply the functional renormalization group. The FRG is a formally exact method, which interpolates between the microscopic action $S$ and the effective action $\Gamma$ via a functional differential equation. The interpolating functional is denoted by $\Gamma_k$, where $k$ is a running momentum scale. $\Gamma_k$ represents an effective action for which fluctuations with momenta $q>k$ have been integrated out, while fluctuations with momenta $q<k$ remain gapped. Taking $\Lambda$ as the ultraviolet cutoff of the theory, e.g. the microscopic lattice spacing, $\Gamma_{\Lambda}=S$ represents the microscopic action, for which no fluctuations have been integrated out. On the other hand,  for $k\rightarrow0$ all fluctuations have been taken into account and $\Gamma_{k=0}=\Gamma$ represents the full effective action. The evolution equation for $\Gamma_k$, smoothly interpolating between these limiting cases, is the Wetterich equation \cite{Wetterich1993}, 
\eq{Wetterich}{
\partial_k\Gamma_k=\frac{1}{2}\text{Tr}\left[\left(\Gamma^{(2)}_k+R_k\right)^{-1}\partial_kR_k\right],
}
which describes the flow of $\Gamma_k$ as a function of the matrix of second order functional derivatives with respect to the fields $\Gamma_k^{(2)}$ and the RG cutoff $R_k$, which we will discuss below. Equation \eqref{Wetterich} represents an exact, non-perturbative expression for the effective action, which we will in the following solve approximately by restricting the form of $\Gamma_k$ to the form of the microscopic action $S$ in Eq.~\eqref{Saddle2}. Furthermore, we perform a derivative expansion of the couplings to leading order in frequency and momentum dependence. This yields 
\eq{Truncation}{
\Gamma_k=\int_X \tilde{\varphi}\left(Z_k\partial_t-D_k\nabla^2+\rho_k\right)\varphi+\mu_k \tilde{\varphi}\varphi\left(\varphi-\tilde{\varphi}\right)-\alpha_k\rho_k\tilde{\varphi}^2.
}
Here, $Z_k, D_k$ and $\rho_k$ are the wave function renormalization, diffusion constant and mass term and $\mu_k$ is the non-linearity. All these parameters depend on $k$ and flow during the RG procedure but have no explicit dependence on the fields $\tilde{\varphi}, \varphi$. Additionally, we introduced the flowing parameter $\alpha_k$, which represents an imbalance between the density and the response field, and explicitly breaks the modified rapidity inversion symmetry \eqref{modsym} for values $\alpha_k\neq1$. Due to the absence of rapidity inversion symmetry in the active phase as argued above, $\alpha_k$ may acquire a non-trivial renormalization group flow. However, the parameter $\alpha_k$ is treated as a common coupling in the theory, which has canonical scaling dimension of zero and does not acquire an anomalous dimension, such that the typical DP scaling $\tilde{\varphi}\sim\varphi$ is preserved within this approach. The coupling $\alpha_k$ does not contribute an additional independent critical exponent and rather contributes to the block of static critical exponents in the stability matrix of the renormalization group flow. We want to stress, that for a cubic truncation of $\Gamma$, only one additional vertex appears in the presence of the background field and thus $\alpha_k$ is the only independent parameter, which can be added to the truncation \eqref{Truncation}. Once this is done, the flow of all remaining couplings is uniquely determined.

One possible way to interpret $\alpha_k$ is to see it as the ratio between the gap and the quadratic noise term. Microscopically this ratio is unity but during the RG flow it will become much larger than one. While the gap and the noise both vanish at the transition, this ratio remains finite even at the transition.
This indicates a strongly increased noise level when the transition from the active to the inactive phase is approached. As can be inferred from the fixed point values of $\alpha_k$, this enhancement of the noise level becomes more an more pronounced, the lower the dimensionality of the system is.
\begin{figure}[t]
  \includegraphics[width=6cm]{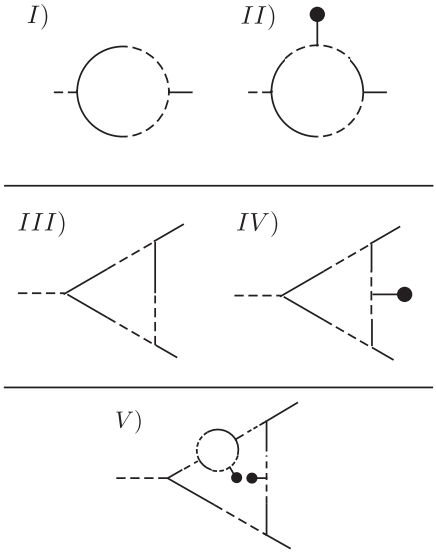}
  \caption{Schematic illustration of the loop diagrams contributing to the renormalization of the retarded propagator (upper row) and the nonlinearity $\mu$ (second row). While the diagrams in the left column are present as well in the inactive phase, the diagrams in the right column emerge only in the active phase. The dot represents a density expectation value proportional to the background field $\rho$. The bottom row ($V$) shows a two-loop diagram generated by a combination of $II)$ and $IV)$ in the presence of the background field. It represents an effective three-loop diagram, proportional to the background field correlation function $\sim \rho^2$.}
  \label{fig:Loops}
\end{figure}
The technical advantage of the present formalism is the effective inclusion of higher order diagrams and vertices via the background field $\rho$. This is elucidated in Fig.~\ref{fig:Loops}, which schematically displays the one-loop correction to the self-energy (upper row) and the nonlinearity (second row). It shows that four-body vertex corrections $\sim \varphi^2\tilde{\varphi}^2, \tilde{\varphi}\varphi(\varphi-\tilde{\varphi})^2$ are partially incorporated in the effective three-body vertex correction (e.g. $IV$ in Fig.~\ref{fig:Loops} is proportional to the effective four-body term $\varphi^2\tilde{\varphi}\rho$), which improves the results compared to a cubic truncation in the inactive phase. Furthermore, two-loop diagrams consisting of $II$ and $IV$ effectively consist of three-loop diagrams with a single, closed density line, as displayed in the bottom row via $V$. As a consequence, Eqs.~\eqref{Wetterich}, \eqref{Truncation} represent an effective  higher loop approach, which adds an effective field dependence to the renormalization of the wave function renormalization and the diffusion constant, i.e. $D_k\rightarrow D_k(\rho_k), Z_k\rightarrow Z_{k}(\rho_k)$. 
This represents a very important improvement, since for a static truncation (i.e. no frequency dependence of $Z_k, D_k, \rho_k, \mu_k$), effective higher order loop corrections improve the estimate for the anomalous dimension of the diffusion constant and the wave function renormalization drastically. This is for instance well established for RG approaches to the $O(n)$ model in the symmetry broken phase \cite{Berges2002}, where the background field method provides a momentum dependent one-loop contribution with two external condensate lines capturing the leading divergence of the (two-loop) setting sun diagram obtained, e.g., in second order $\epsilon$ expansion. Indeed, the present implementation shows a significant improvement for the estimate of these exponents especially in lower dimensions, which we attribute to the effective higher loop computation. 


The truncation \eqref{Truncation} can be straightforwardly extended to consider higher order contributions such as quartic vertices as well. Then the imbalance between density and response field must be taken into account also on the cubic level, yielding the correction $\delta\Gamma_k$ to \eqref{Truncation}
\eq{Quart}{
\delta\Gamma_k=\int_X \lambda_k \tilde{\varphi}^2\varphi^2+\kappa_k \tilde{\varphi}\varphi(\tilde{\varphi}-\varphi)^2-\mu_k(\beta_k-1)\tilde{\varphi}^2\varphi.
}
This might possibly lead to a slight improvement for the estimates of the static exponents $\beta, \nu$ but also leads to a significantly enhanced complexity of the corresponding flow equations compared to the ones discussed in this work. 

\section{Flow equations and universal exponents}\label{sec:Exp}
We proceed by determining the flow equations for the parameters in the effective action \eqref{Truncation} and evaluate the corresponding critical exponents at the fixed point. In order to do so, we specify the cutoff function $R_k$ to be the optimized FRG cutoff \cite{Litim2001} with matrix structure
\eq{Flowequations}{
R_k=D_k(k^2-q^2)\theta(k^2-q^2)\sigma_x,
}
where $q$ is the external momentum of the fields $\tilde{\varphi}, \varphi$ and $\sigma_x$ is the Pauli matrix.
It acts in the advanced/retarded sector ($\sim \tilde{\varphi}\varphi$) of the action $\Gamma_k$ and replaces spatial fluctuations below the momentum scale $k$ by a constant gap $D_kk^2$, while it vanishes at momenta $q>k$. 

The flow equations for the running couplings are obtained by taking the corresponding field functional derivatives of the Wetterich equation \eqref{Wetterich}. This yields
\begin{eqnarray}
\partial_k\mu_k=\frac{1}{2}\left.\frac{\delta^3(\partial_k\Gamma_k)}{\delta\tilde{\varphi}_0\delta\varphi_0\delta\varphi_0}\right|_{\tilde{\varphi}=\varphi=0}, \ \ \ \partial_k(\rho_k\alpha_k)=\left.-\frac{1}{2}\frac{\delta^2(\partial_k\Gamma_k)}{\delta\tilde{\varphi}_0\delta\tilde{\varphi}_0}\right|_{\tilde{\varphi}=\varphi=0},\ \ \label{Flowequations2}
\end{eqnarray}
where $\tilde{\varphi}_0, \varphi_0$ are shorthand for $\tilde{\varphi}_{\omega=0, q=0}, \varphi_{\omega=0, q=0}$ in frequency and momentum space. The flow of the inverse retarded propagator is 
\eq{Flowequations3}{
\partial_k G^{-1}_k(\omega,q)=\left. \frac{\delta^2\Gamma_k}{\delta\tilde{\varphi}_{-\omega,-q}\delta\varphi_{\omega,q}}\right|_{\tilde{\varphi}=\varphi=0}
}
and since $G^{-1}_k(\omega,q)=-iZ_k\omega+D_kq^2+\rho_k$, it follows
\begin{eqnarray}
\partial_kZ_k&=&\left.i\partial_\omega\partial_kG_k^{-1}(\omega,q)\right|_{\omega=q=0},\\
 \partial_kD_k&=&\left.\partial_{q^2}\partial_kG^{-1}(\omega,q)\right|_{\omega=q=0},\label{MoLem}\\
\partial_k\rho_k&=&\partial_kG^{-1}_k(\omega=0,q=0),\label{Flowequations6}
\end{eqnarray}
where the evaluation of Eq.~\eqref{MoLem} incorporates some technical subtleties \cite{Morris1994}. 
Finally, we combine Eqs.~\eqref{Flowequations2}, \eqref{Flowequations6} to find
\eq{Flowequations7}{
\partial_k\alpha_k=-\frac{1}{\rho_k}\left(\frac{1}{2}\frac{\delta^2(\partial_k\Gamma_k)}{\delta\tilde{\varphi}_0\delta\tilde{\varphi}_0}+\alpha_k\frac{\delta^2(\partial_k\Gamma_k)}{\delta\tilde{\varphi}_0\delta\varphi_0}\right)_{\tilde{\varphi}=\varphi=0}.
}
In order to find dimensionless expressions for the renormalization group flow, we introduce the dynamical critical exponent $\eta_Z$ and the anomalous dimension $\eta_D$, which are defined as
\eq{Flowequations8}{
\eta_Z=-\frac{k}{Z_k}\partial_kZ_k, \ \ \ \eta_D=-\frac{k}{D_k}\partial_kD_k.
}
Furthermore, we define the dimensionless gap, nonlinearity and diffusion constant according to the canonical scaling via $\tilde{\rho}_k=\rho_k k^{-2+\eta_D}$, $\tilde{\mu}_k=\mu_kZ_k^{-1/2}k^{\frac{d}{2}-2+\eta_D}$ and $\tilde{D}_{k}=D_kk^{\eta_D}$. We thus see that $\rho_k \sim k^2$ on the canonical scaling level, removing a Markovian noise level and restoring the conventional scaling according to the rapidity inversion symmetry. After rescaling time $t\rightarrow tZ_kk^{\eta_D}$, the corresponding flow equations in $d$ dimensions are
\begin{eqnarray}
k\partial_k\tilde{\rho}_k&=&\left(\eta_D-2-\tfrac{2(2+d-\eta_D)(\tilde{D}_k+(1+2\alpha_k)\tilde{\rho}_k)}{\tilde{\rho}_kd(2+d)(\tilde{D}_k+\tilde{\rho}_k)^3}C_d\tilde{D}_k\tilde{\mu}_k^2\right)\tilde{\rho}_k,\label{Flow1}\\
k\partial_k\tilde{\mu}_k&=&\left(\eta_D\hspace{-0.1cm}+\hspace{-0.1cm}\tfrac{\eta_Z-4+d}{2}\hspace{-0.1cm}+ \hspace{-0.1cm}\tfrac{4(2+d-\eta_D)(2\tilde{D}_k+(2+3\alpha_k)\tilde{\rho}_k)}{d(d+2)(\tilde{D}_k+\tilde{\rho}_k)^4}C_d\tilde{D}_k\tilde{\mu}^2_k\right)\tilde{\mu}_k, \ \ \ \ \ \ \ \ \ \ \ \\
k\partial_k\alpha_k&=&-\tfrac{(2+d-\eta_D)(4\tilde{D}_k^2+8\tilde{D}_k(2+\alpha_k)\tilde{\rho}_k+(12+11\alpha_k)\tilde{\rho}_k^2)}{2d(2+d)\tilde{\rho}_k(\tilde{D}_k+\tilde{\rho}_k)^4}C_d\tilde{D}_k\tilde{\mu}_k^2\alpha_k,\label{Flow3}
\end{eqnarray}
and $C_d$ is the surface of the $d$-dimensional unit sphere.

We are interested in the critical point of the active to inactive phase transition of the system, which features a scale invariant effective action $\Gamma_k$, i.e. a vanishing flow of the dimensionless couplings $\tilde{\mu}_k, \tilde{\rho}_k$, as well as $\alpha_k$, expressed by the conditions
\eq{Flowequations12}{
\partial_k\tilde{\rho}_k=\partial_k\tilde{\mu}_k=\partial_k\alpha_k=0.} 
In the following we denote the fixed point values of the couplings with a star. The trivial solution of Eq.~\eqref{Flowequations12} is $\tilde{\mu}^*=\tilde{\rho}^*=0$ and $\alpha^*$ arbitrary. This marks the Gaussian fixed point, which is stable in dimensions $d\ge 4$ and leads to the mean-field scaling behavior discussed in the previous section. For $d<4$, this fixed point becomes unstable and the desired fixed point, which determines the universal exponents, is the Wilson-Fisher fixed point with $\tilde{\mu}^*,\tilde{\rho}^*, \alpha^*\neq0$. 

Equations \eqref{Flow1}-\eqref{Flow3} can be solved analytically, yielding a set of possible fixed point values $(\tilde{\rho}^*,\tilde{\mu}^*,\alpha^*)$. For dimensions $d<4$ there exists only a single combination, which corresponds to a physical and stable fixed point (i.e. a fixed point with only a single relevant direction in the $\rho-\mu$ sector of the stability matrix). The explicit formulas for the fixed point values are rather complicated and we will not discuss them here. However, one can perform an expansion of the expressions in the parameter $\epsilon=4-d$ in order to compare it with the known values from an $\epsilon$-expansion based renormalization group approach \cite{DP_Hinrichsen,Janssen1981,Janssen2001,Bronzan1974}. We find
\begin{eqnarray}
\tilde{\rho}^*&=&-\tilde{D} \left(\frac{\epsilon}{12}-\frac{\epsilon^2}{288}+\mathcal{O}(\epsilon^3)\right),\label{e1}\\
(\tilde{\mu}^*)^2&=&\frac{\tilde{D}^2}{C_d}\left(\frac{\epsilon}{3}-\frac{31\epsilon^2}{108}+\mathcal{O}(\epsilon^3)\right),\label{e2}\\
\alpha^*&=&-3-\frac{5\epsilon}{4}-\frac{107\epsilon^2}{192}+\mathcal{O}(\epsilon^3).\label{e3}
\end{eqnarray}
Although the present formalism is non-perturbative, it should agree with the $\epsilon$-expansion results up to first order, as only bare vertices and propagators contribute to first order in $\epsilon$, such that the FRG and $\epsilon$-expansion coincide. Equations \eqref{e1}, \eqref{e2} show that this is the case up to a redefinition of $D\rightarrow 2D$ in Ref.~\cite{DP_Hinrichsen}. However, one also realizes that already at second order in $\epsilon$ the difference of the FRG result from the $\epsilon$-expansion is significant. For the parameter $\alpha^*$, which has no counterpart in the $\epsilon$-expansion, one observes a strong dependence on the parameter $\epsilon$, i.e. on the dimensionality, indicating that the modification of the critical exponents in low dimensions is pronounced compared to a bare one-loop approach with $\alpha^*=0$. The fact that $\alpha_k$ remains finite even for $\epsilon=0$ indicates that the ratio between the noise $\sim \alpha_k\rho_k$ and the gap $\sim\rho_k$ in Eq.~\eqref{Truncation} remains finite even in $d=4$ while the individual terms vanish equally fast for $d\rightarrow4$, as expected. We would like to add here, that the series expansion of $\alpha$ in Eq.~\eqref{e3} is only convergent for $\epsilon\le 2$ and the corresponding fixed point becomes unstable in $d=1$. While Eqs~\eqref{e1}, \eqref{e2} remain valid at the stable fixed point in $d=1$, $\alpha^*$ is strongly modified.\\

\begin{table}[H]
\begin{center}
\begin{tabular}{|C{2cm} |C{2cm} | C{2cm}| C{2cm}|}
  \hline
  \rule{0mm}{3ex}
Exponent & $d=3$&$d=2$& $d=1$\\
  \hline\hline
\rule{0mm}{3ex} $\beta_{\text{FRG}}$& $0.846$&$0.640$ &$0.251$\\ \rule{0mm}{3ex} 
$\beta_{\text{Num}}$&$0.81(1)$ &$0.584(4)$ &$0.276486(8)$\\ 
\hline
\rule{0mm}{3ex} $\nu_{\text{FRG}}$ &$0.595$ &$0.762$ &$1.011$\\ \rule{0mm}{3ex} 
$\nu_{\text{Num}}$ &$0.581(5)$ &$0.734(4)$ &$1.096854(4)$\\
\hline
\rule{0mm}{3ex} $z_{\text{FRG}}$& $1.903$&$1.777$ &$1.609$  \\ \rule{0mm}{3ex} 
$z_{\text{Num}}$&$1.90(1)$ &$1.76(3)$ &$1.580745(10)$ \\
  \hline
\end{tabular}
\caption{Numerical values of the critical exponents obtained from the present approach (subscript FRG) compared to exact results from entirely numerical simulations (subscript Num) for dimensions $d=3$ \cite{Jensen1992}, $ d=2$ \cite{Voigt1997}, $ d=1$ \cite{Jensen1999}. For the static exponents, the deviation of the FRG results is between five and ten percent, while the deviation of the dynamic exponents is remarkably small and below two percent in all dimensions. For completeness, we also mention the critical exponents obtained by $\epsilon$-expansion \cite{Janssen1981}: $z=2-\frac{\epsilon}{12}\left(1+0.30505\epsilon\right),$ $\nu=\frac{1}{2}+\frac{\epsilon}{16}\left(1+0.3376\epsilon\right), \beta=1-\frac{\epsilon}{6}\left(1+0.0677\epsilon\right)
$.}
\label{table:scaling}
\end{center}
\end{table}

The critical exponents for the directed percolation transition can be obtained from the fixed point values $(\tilde{\rho}^*, \tilde{\mu}^*, \alpha^*)$ according to the following relations \cite{Canet2004, Tauber2005} \footnote{The expression $\left. \partial_{\tilde{\rho}}\partial_k\tilde{\rho}\right|_{\alpha_k\rho_k}$ means that products $\alpha_k\rho_k$ are kept fixed when performing the derivative since $\alpha_k\rho_k$ is independent from the gap.}
\begin{eqnarray}
z&=&2+\eta_Z-\eta_D=2-\frac{\epsilon}{12}-\frac{5\epsilon^2}{432}+\mathcal{O}(\epsilon^3),\label{eqz}\\
\nu&=&-\left(\left. k\partial_{\tilde{\rho}}\partial_k\tilde{\rho}\right|_{\alpha_k\tilde{\rho}_k}\right)^{-1}=\frac{1}{2}+\frac{\epsilon}{16}+\frac{125\epsilon^2}{3456}+\mathcal{O}(\epsilon^3),\ \ \ \ \ \ \ \ \ \ \\
\beta&=&\frac{d+\eta_Z}{2}\nu=1-\frac{\epsilon}{6}+\frac{31\epsilon^2}{864}+\mathcal{O}(\epsilon^3),\label{eqb}
\end{eqnarray}
which we again expanded up to second order in $\epsilon$ in order to compare it to known results from the $\epsilon$-expansion. Again, the first order expansion in $\epsilon$ coincides with the perturbative RG calculations, while the second order term shows strong deviations from this method. Compared to a perturbative expansion, the $\epsilon$-series in our approach is fastly converging to the FRG results. The numerical values for the critical exponents obtained from our approach (without expansion in $\epsilon$) are shown in Table \ref{table:scaling} and compared with exact numerical results \cite{Jensen1992, Jensen1999, Voigt1997}. 
The agreement of the FRG results with the exact results is remarkably good, even in low dimensions. Especially the exponents $\eta_D, \eta_Z$ are determined very accurately, which leads to good estimates of the exponents $z$ and $\beta$, which depend on the dynamical critical exponent and the anomalous dimension via the scaling relations \eqref{eqz}, \eqref{eqb}.

We conclude this section with a comment on physical meaning of the parameter $\alpha_k$. This parameter represents the ratio and therefore the imbalance between the gap and the quadratic noise level, which is generated during the RG flow. As such it leads to a modification of the density-density correlation function 
\eq{Lastequation}{
C_k(\omega,q)=\langle \varphi_{-\omega,-q}\varphi_{\omega,q}\rangle=\frac{\alpha_k\rho_k}{Z_k^2\omega^2+(D_kq^2+\rho_k)^2}+\rho_k^2\delta(\omega)\delta(q).
}
This correlation function consists of a static contribution ${\sim \rho_k^2}$ and a dynamic correction, which is modified by spatio-temporal fluctuations as well as the noise kernel $\alpha_k\rho_k$. Within the present approach, the noise is generally enhanced $\alpha_k>1$ due to interactions as compared to the bare prediction $\alpha_k=1$ in the absence of a renormalization. Interpreting the present results in view of this brief discussion, we conclude that the enhancement of the effective noise level is an important feature at the active to inactive transition. Allowing the system to explicitly build up this imbalance during the RG flow results in an improved estimate for the critical exponents in the system.
One should, however, not be confused by the negative value of $\alpha^*$ at the fixed point, since it goes in hand with a negative gap $\rho^*$, such that $\rho^*\alpha^*>0$. In principle, negative values of the product $\alpha_k\rho_k$ are possible, leading to suppressed density correlations for the corresponding parameter regime. However, the determination of the correlations away from the fixed point is beyond the scope of the present approach.

\section{Conclusion}\label{sec:Conc}
We have presented a background field functional renormalization group approach for the directed percolation transition starting from the active, i.e. finite density, phase. This method represents a non-perturbative, effective higher loop computation, which interpolates between the first order $\epsilon$-expansion and the exact results obtained from numerical Monte-Carlo studies. Due to the finite truncation of the effective action in the active phase, the implemented renormalization group scheme explicitly breaks the rapidity inversion symmetry during the RG-flow; it, however, restores the correct scaling behavior of the fields at the fixed point, and is therefore well suited to determine universal scaling behavior. The gain of this symmetry breaking approach is, on the technical side, the effective inclusion of a field dependence of the wave function renormalization and diffusion constant due to higher order loop contributions compared to previous functional renormalization group schemes. In terms of physics, the present approach introduces the enhancement of the noise level compared to the microscopic theory, which on the basis of our results seems to be an important contribution for absorbing state phase transitions.
This leads to a remarkable improvement in the estimates for the exponent $z$, similar to RG-approaches for $O(n)$-models in the symmetry broken phase. This method represents a simple, analytically tractable, yet reasonably accurate tool to investigate the universal properties of active to inactive state phase transitions, with a spectrum of applications, e.g. non-equilibrium and absorbing state phase transitions beyond the directed percolation universality class.

\begin{acknowledgments}
We thank Dietrich Roscher for comments on the manuscript and enlightening discussions.
We acknowledge funding by the German
Research Foundation (DFG) through the Institutional
Strategy of the University of Cologne within the German
Excellence Initiative (ZUK 81) and the European Research Council (ERC) under the European UnionÕs
Horizon 2020 research and innovation programme (grant agreement No
647434).
\end{acknowledgments}

\bibliography{biblio}

\end{document}